\documentclass[aps,twocolumn,showpacs]{revtex4}
\usepackage{amssymb}
\usepackage{graphicx}
\usepackage{dcolumn}
\usepackage{bm}

\begin{document}

\title{Statistical investigation and thermal properties for a 1-D impact system 
with dissipation}

\author{Gabriel D\'iaz I.$^1$, Andr\'e L.\ P.\ Livorati$^{1}$ and Edson D.\ Leonel$^{1}$}

\affiliation{$^1$ Departamento de F\'isica -- UNESP -- Univ Estadual Paulista -- Av. 24A, 1515 - Bela Vista -
13506-900 - Rio Claro - SP - Brazil.}


\begin{abstract}
The behavior of the average velocity, its deviation and average squared velocity 
are characterized using three techniques for a 1-D dissipative impact system. 
The system -- a particle, or an ensemble of non interacting particles, moving in 
a constant gravitation field and colliding with a varying platform -- is 
described by a nonlinear mapping. The average squared velocity allows to 
describe the temperature for an ensemble of particles as a function of the 
parameters using: (i) straightforward numerical simulations; (ii) analytically 
from the dynamical equations; (iii) using the probability distribution 
function. Comparing analytical and numerical results for the three techniques, 
one can check the robustness of the developed formalism, where we are able 
to estimate numerical values for the statistical variables, without doing extensive
numerical simulations. Also, extension to other dynamical systems is immediate,
including time dependent billiards.
\end{abstract}

\maketitle

\section{Introduction}
\label{sec1}

In the last decades, modeling of dynamical systems, especially low-dimensional 
ones, becomes one of the most challenging areas of interest among 
mathematicians, physicists \cite{ref1,ref2,ref3} and many other sciences. 
Depending on both the initial conditions as well as control parameters, such 
dynamical systems may present a very rich and hence complex dynamics, therefore 
leading to a variety of nonlinear phenomena. The dynamics can be considered 
either in the dissipative or non-dissipative regime \cite{ref4,ref5,ref6} 
yielding into new approaches, new formalisms therefore moving forward the 
progress of nonlinear science.

Since the so called Boltzmann ergodic theory \cite{ref5,ref6}, the assembly 
between statistical mechanics and thermodynamics has produced remarkable 
advances in the area leading also to progress in experimental and observational 
studies \cite{ref7,ref8,ref9,ref10,ref11}. Indeed, statistical tools can be 
used for a complete analysis of the dynamical behavior of such type of systems. 
Depending on the control parameters, phase transitions and abrupt changes in 
the phase space can be observed in time as well as in parameter space 
\cite{ref6} while many results can be described by using scaling laws approach 
\cite{ref12}. In this paper we revisit the 1-D impact system aiming to obtain 
and describe the behavior of average properties in the chaotic dynamics 
focusing in the stationary state, {\it id est}, for very long time, where transient 
effects are not influencing the dynamics anymore. 
Analytical expressions will be presented in order to calculate statistical properties 
for the average velocity, its deviation and the average squared velocity, when
these variables reach the stationary state. The developed formalism, allows us
to obtain the numerical values for these variables, without doing the numerical
simulations. We will show a remarkable agreement between numerical simulations
and theoretical analysis considering either statistical and thermal variables,
giving so robustness, to the developed theory.

The impact system is described by a free particle, or an ensemble of non 
interacting particles, moving under the presence of a constant gravitational 
field and experiencing collisions with a heavily vibrating platform 
\cite{ref13,ref14}. For elastic collisions, the dynamics leads to a mixed phase 
space, described in velocity and time, and two main properties are observed 
according to the control parameter range. If the parameter is smaller than a 
critical one, invariant spanning curves, also called as invariant tori, are 
present in the phase space hence limiting the velocity of the particle in a 
chaotic diffusion for certain portions of the phase space. On the other hand, 
for a parameter larger than the critical one, invariant spanning curves are not 
present anymore and unlimited diffusion in velocity, for specific ranges of 
initial conditions, can be observed. The scenario is totally different when 
inelastic collisions are considered. In this case, dissipation is in course, 
hence contracting area in the phase space, therefore leading to the existence 
of attractors. For strong dissipation and control parameter beyond the critical 
one, attractors are most periodic. For weak dissipation and large control 
parameter, chaotic attractors, characterized by a positive Lyapunov exponent 
\cite{eckmann_ruelle}, dominate over the phase space. Giving the 
attractors are far away from the infinity (in velocity axis), dissipation has 
proved to be a powerful way of suppress unlimited diffusion. Because of limited 
diffusion in phase space, the behavior and properties for both average 
velocity, 
average squared velocity or the deviation around the average velocity, known 
also as roughness, are the following. They grow to start with from a low 
initial velocity value and, eventually, they bend towards a stationary state 
\cite{ref15,ref16} at very long time. The scenario is scaling invariant with 
respect to the control parameters and number of collisions with the moving 
platform. By the use of equipartition theorem, the steady state, obtained in 
the 
asymptotic state, can be used to make a connection with the thermal equilibrium 
of the system \cite{ref16}. Therefore in the present paper, we evaluate 
numerically, for long time series, the behavior of: (i) the average velocity; 
(ii) the averaged squared velocity; and (iii) the deviation around the average 
velocity, both for the dissipative impact system. We then compare the numerical 
results with analytical expressions at the equilibrium, obtained via 
statistical 
and thermodynamics analysis by using the dynamical equations \cite{ref16}. A 
comparison between the results obtained using numerical simulation and 
theoretical investigation is remarkable, hence giving robustness to the 
connection between statistical mechanics, thermodynamics and the modeling of 
dynamical systems. It also improves the theoretical formalism that 
can be extended to other different types of systems including the time 
dependent billiards.

The paper is organized as follows: in Sec. \ref{sec2} we describe the dynamics 
of the impact system and some of its properties. Section \ref{sec3} is devoted 
to the discussion of the numerical investigation. The results using the 
dynamical equations and connection with the thermodynamics in the stationary 
state and the discussions of the results are presented in Sec. \ref{sec4}. 
Finally, Sec. \ref{sec5} brings some final remarks and conclusions.

\section{The model, the mapping and some statistical properties}
\label{sec2}

The model we consider consists of a particle\footnote{Or an ensemble of non 
interacting particles.} of mass $\mu$ moving under the action of a 
gravitational 
field and experiences collisions with a heavy periodically moving 
wall. This model is also referred to as a bouncer or bouncing ball model. It 
backs to Pustylnikov \cite{ref17} and has been studied for many years 
\cite{ref18,ref19,ref20,ref21}, with several applications in different areas of 
research such as vibration waves in a nanometric-sized mechanical contact 
system \cite{ref22}, granular materials \cite{ref23,ref24,ref25,ref26,ref27}, 
dynamic stability in human performance \cite{ref28}, mechanical vibrations 
\cite{ref29,ref30,ref31}, chaos control \cite{ref32,ref33}, crises between attractors \cite{crises},
among many others.

As usual, the dynamics of the system is described by a two-dimensional, 
non-linear discrete mapping for the variables velocity of the particle $v$ and 
time $t$ (will be measured latter on as function of the phase of moving wall) 
immediately after a $n^{th}$ collision of the particle with the moving wall. 
See Ref. \cite{ref34} for an analysis as function of the time. The 
investigations are made based on two main versions of the model: {\it(i)} 
complete, which takes into account the whole movement of the vibrating platform; 
and {\it(ii)} a static wall approximation. In this version, the nonlinear 
mapping assumes the wall is static but that, as soon as the particle hits it, 
there is an exchange of energy as if the wall were moving. This is then a 
simplified version and shows to be a very convenient way to find out analytical 
results in the model where transcendental equations do not need to be solved, as 
they have to be in the complete version. The two versions can be used either to 
investigate  non-dissipative \cite{ref35} and dissipative dynamics 
\cite{ref13,ref14}. Dissipation here is introduced by using a restitution 
coefficient $\gamma\in[0,1]$ upon collision. For $\gamma=1$ the system is non 
dissipative albeit area contraction in the phase is observed for $\gamma<1$.

To construct the mapping, we consider the motion of the platform is described 
by $y_w(t_n)=\varepsilon\cos{w t_n}$, where $\varepsilon$ and $w$ are, 
respectively, the amplitude and frequency of oscillation. Moreover, we assume 
that at the instant $t_n$, the position of the particle is the same as the 
position of the moving wall, hence $y_p(t_n)=y_w(t_n)$ and with velocity 
$V_n>0$. The mapping then gives the evolution of the states from $(V_n,t_n)$ to 
$(V_{n+1},t_{n+1})$, from $(V_{n+1},t_{n+1})$ to $(V_{n+2},t_{n+2})$ and so on. 
To obtain the analytical expressions of the mapping, we have to take into 
account the time of flight the particle moves without colliding with the wall 
and, from it, determine the velocity of the moving wall upon collision. From 
conservation of momentum law we obtain the velocity of the particle after 
collision. We have indeed four control parameters $g$, $\varepsilon$, $w$ and 
$\gamma$ and not all of them are relevant for the dynamics. Defining 
dimensionless and hence more convenient variables we have $V_n=v_n w/g$ 
(dimensionless velocity) and $\epsilon=\varepsilon w^2/g$, which is the ratio 
between accelerations of the vibrating platform and the gravitational field. We 
may also measure the time in terms of the number of oscillations of the moving 
wall $\phi_n=w t_n$. Using this set of new variables, the mapping is written as
\begin{equation}
T_c:\left\{\begin{array}{ll}
V_{n+1}=-\gamma({V_n^*}-{\phi_c})-(1+\gamma)\epsilon\sin(\phi_{n+1})\\
\phi_{n+1}=[\phi_n+\Delta T_n]~~{\rm mod (2\pi)}\\
\end{array}
\right.,
\label{eq1}
\end{equation}
where the sub-index $c$ stands for the complete version of the model. The 
expressions for $V_n^*$ and $\Delta T_n$ depend on what kind of collision
happens. For the case of multiple collisions, those the particle experiences 
without leaving the collision zone (a region in space where the moving wall is 
allowed to move), the corresponding expressions are $V_n^*=V_n$ and $\Delta 
T_n=\phi_c$ where $\phi_c$ is obtained from the condition that matches the same 
position for the particle and the moving wall. It leads to the following 
transcendental equation that must be solved numerically
\begin{equation}
G(\phi_c)=\epsilon\cos(\phi_n+\phi_c)-\epsilon\cos(\phi_n)-V_n\phi_c+{{
1}\over{2}}\phi_c^2~.
\label{eq2}
\end{equation}

If the particle leaves the collision zone, than indirect collisions are 
observed. The expressions for the velocity and phase are 
$V_n^*=-\sqrt{V_n^2+2\epsilon(\cos(\phi_n)-1)}$ 
and $\Delta T_n=\phi_u+\phi_d+\phi_c$ with $\phi_u=V_n$ denoting the time spent 
by the particle in the upward direction up to reach the null velocity while the 
expression 
$\phi_d=\sqrt{V_n^2+2\epsilon(\cos(\phi_n)-1)}$ corresponds to the time the 
particle spends from the place where it had zero velocity to the entrance of 
the collision zone. Finally the term $\phi_c$ has to be obtained numerically 
from the equation $F(\phi_c)=0$ where
\begin{equation}
F(\phi_c)=\epsilon\cos(\phi_n+\phi_u+\phi_d+\phi_c)-\epsilon-V_n^*
\phi_c+{{1}\over{2}}\phi_c^2~.
\label{eq3}
\end{equation}    

For the static wall approximation \cite{karlis}, where no transcendental 
equations must be solved, the mapping has the form
\begin{equation}
T_{swa}:\left\{\begin{array}{ll}
V_{n+1}= |(\gamma V_n)-(1+\gamma)\epsilon\sin(\phi_{n+1})|\\
\phi_{n+1}=[\phi_n+2V_n]~~{\rm mod (2\pi)}\\
\end{array}
\right..
\label{eq4}
\end{equation}
The static wall approximation ($swa$), as quoted in the sub-index of mapping 
(\ref{eq4}) is convenient to avoid solving transcendental equations. However, 
it inherently introduce a new problem that must be taken into account prior 
evolve the dynamical equations. In the complete version, after a collision with 
the moving wall, the particle, in specific cases and under certain conditions, 
can keep moving downward with negative velocity. Of course if would lead to a 
successive collision in such a version of the model. In the static wall 
approximation, this type of collision is not allowed and a negative velocity 
would necessarily produce a non physical situation. To avoid this unphysical 
case, the modulus function is introduced and prevents the particle of the 
possibility of moving beyond the wall. When such a condition happens, the 
particle is just re-injected back into the dynamics with the same velocity 
before the collision, however in the upward direction. If the velocity is 
positive after a collision, the modulus function does not affect nothing the 
equation.

\begin{figure*}[t]
\begin{center}
\centerline{\includegraphics[width=14cm,height=12cm]{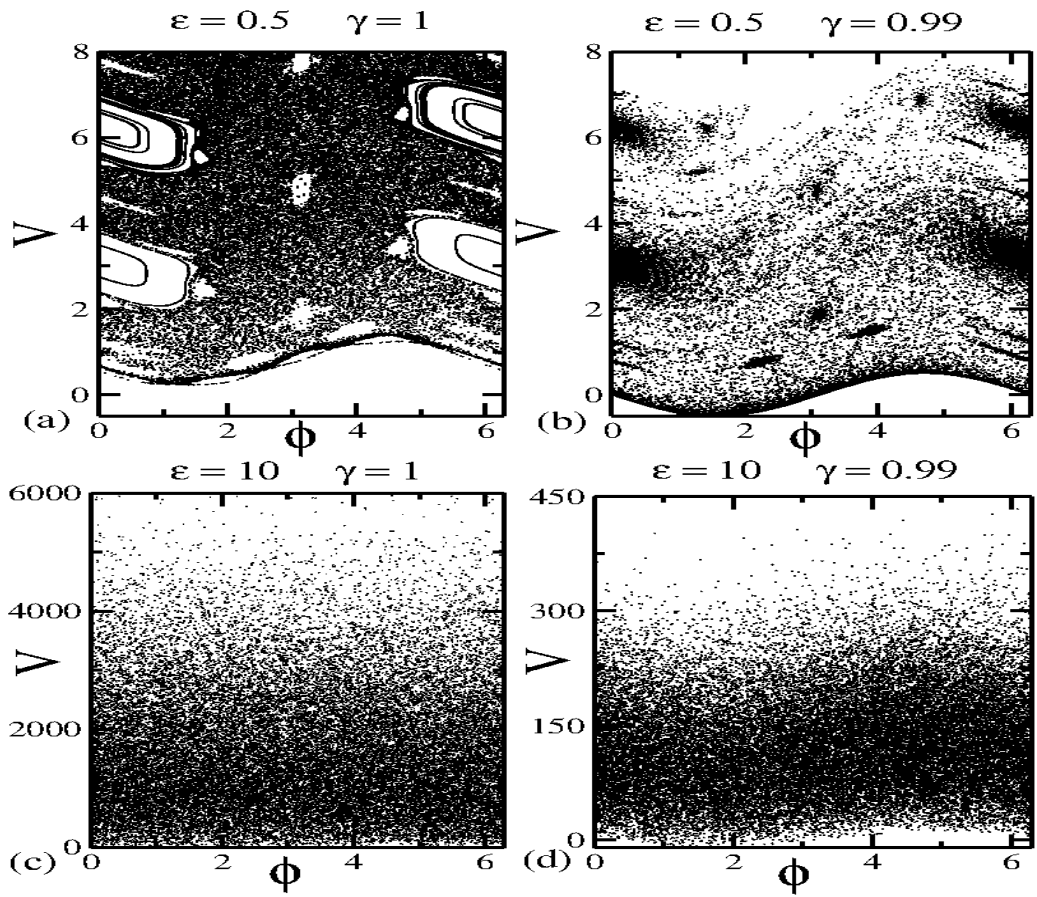}}
\end{center}
\caption{\it{Snapshots of $(V,\phi)$ for the impact system considering either 
non-dissipative and dissipative dynamics. The control parameters used were: (a) 
$\epsilon=0.5$ and $\gamma=1$; (b) $\epsilon=0.5$ and $\gamma=0.99$; (c) 
$\epsilon=10$ and $\gamma=1$; and (d) $\epsilon=10$ and $\gamma=0.99$.}}
\label{fig1}
\end{figure*}

Figure \ref{fig1} shows the phase space considering both 
non-dissipative and dissipative dynamics for the complete model. We used $100$ 
different initial conditions iterated up to $10^4$ collisions. Figure 
\ref{fig1}(a) shows the phase space for $\gamma=1$ and $\epsilon=0.5$. Easily 
observed and typical of Hamiltonian systems is the mixed dynamics scenario. It 
contains, indeed, stability islands and chaotic seas. Because of the absence of 
invariant tori -- invariant spanning curves limiting the size of the 
chaotic sea -- unlimited diffusion in velocity is observed. This phenomenon is 
known also as Fermi Acceleration (FA) \cite{ref36} can be slowed down by the 
presence of stickiness \cite{ref35}. In this case, a chaotic orbit may passes 
nearby a stability island and be trapped there around it for a 
finite\footnote{Sometimes very long time.} time \cite{ref3,ref4}. Opposite 
to trapping, the so called accelerating modes, produced by resonances, can 
affect globally the dynamics \cite{ref37} leading to a fast acceleration.

Dissipation, introduced by inelastic collisions, however destroys the mixed 
structure of the phase space. As shown in 
Fig. \ref{fig1}(b) for $\gamma=0.99$ and $\epsilon=0.5$, the blurred points, 
suggesting a chaotic attractor, represent nothing more than transient orbits, 
which shall settle down at asymptotic fixed points (sinks) for a sufficiently 
long time. Figure \ref{fig1}(c) was constructed using $\epsilon=10$ and 
$\gamma=1$. The mixed structure is not observed at this scale and only chaotic 
orbits, diffusing unlimitedly are observed. Finally, Fig. \ref{fig1}(d) was 
obtained for $\epsilon=10$ and $\gamma=0.99$. The unlimited diffusion was 
replaced by a chaotic attractor, which has a limited range. This suppression 
was indeed expected since the determinant of the Jacobian matrix is written as 
\begin{equation}
{\rm 
Det}J=\gamma^2{{V_n+\epsilon\sin(\phi_n)}\over{V_{n+1}+\epsilon\sin(\phi_{n+1})}
}.
\end{equation}
This result confirms that the introduction of dissipation can be considered as 
a 
powerful mechanism to suppress Fermi acceleration \cite{ref13,ref14}.

\section{Statistical and numerical results}
\label{sec3}

Given the expressions of the mapping are already known, in this section, we 
describe the results obtained by numerical simulations. We focus particularly 
on the statistical analysis for the velocity of the particle. As it is already 
known \cite{ref14,ref15,ref16}, for large $\epsilon$ and in the presence of 
small dissipation, {\it id est},  $\epsilon>10$ and $\gamma>0.99$, the dynamics starting from either low or high velocity 
settles down at a stationary state for enough long time. The plateau of 
a saturation can be obtained from different ways: (i) imposing fixed point 
condition in the first equation of mappings (\ref{eq1}) and (\ref{eq4}), after 
averaging them in an ensemble of phase $\phi\in[0,2\pi]$; (ii) transforming the 
equation of the velocity in the discrete mapping into a differential equation 
and solve it using an ensemble of different initial phases $\theta\in[0,2\pi]$; 
(iii) doing numerical simulations and considering long time dynamics.

Because we have the dynamical equations of the mappings, different statistical 
investigations can be made using different types of averages. An observable 
which is immediate is the average velocity measured along the orbit. It is 
written as
\begin{equation}
V_i(n,\epsilon,\gamma)={1\over n} {\sum_{j=1}^n} V_{j}~.
\label{eq5} 
\end{equation}
We can use Eq. (\ref{eq5}) and average it over an ensemble of different initial 
conditions, hence leading to
\begin{equation}
\langle{V}\rangle={1\over M} {\sum_{i=1}^M} V_i(n,\epsilon,\gamma)~,
\label{eq6} 
\end{equation}
where $M$ represents an ensemble of initial conditions. For instance, the 
initial velocity is assumed constant and $M$ different phases uniformly 
distributed in the range $\phi\in[0,2\pi]$ are considered. The root mean square 
velocity is obtained as
\begin{equation}
V_{rms}=\sqrt{\langle{V^2}\rangle}~.
\label{eq7}
\end{equation}
The procedure is the same as running Eqs. (\ref{eq5}) and (\ref{eq6}) but using 
$V^2$ rather than $V$. Finally, the deviation around the average velocity, 
$\omega$, see \cite{ref12} for instance, is obtained from
\begin{equation}
\omega=\sqrt{\langle V^2\rangle-{\langle V \rangle}^2}~.
\label{eq8}
\end{equation}

As it is known, for large $\epsilon$, unlimited diffusion in velocity can be 
observed. Because of the dissipation, the unlimited diffusion is not allowed 
anymore. The average dynamics, no mater the initial velocity, will converge to 
an asymptotic state for large time. If the initial condition is large, the 
velocity of the particle decreases until reaches the stationary state. It is 
known in the literature for a similar system, that the decay of velocity is 
given by an exponential function \cite{danila1,danila2} and the speed of the 
decay depends on the strength of the dissipation. Stronger the dissipation, 
faster the decay.

In opposite way, starting with a small initial velocity, the dynamics leads the 
average velocity to experience an initial growth as a function of the 
number of collisions of the type $[\epsilon^2n]^{\beta}$. The acceleration 
exponent is $\beta=1/2$, similar to random walk systems, and eventually, the 
growing regime is replaced by a constant plateau. The crossover that marks the 
change from growth to the saturation is described by a power law on 
$(1-\gamma)^{z_2}$, with $z_2=-1$. The average velocity of the particle at 
the stationary regime depends either on the nonlinear parameter as well on the 
dissipation parameter as $\epsilon^{\alpha_1}(1-\gamma)^{\alpha_2}$ where 
$\alpha_1=1$ and $\alpha_2=-1/2$.

When the initial velocity is neither small or large, say below the saturation 
regime, an additional crossover time is observed in the curves 
\cite{leonel100,leonel101}. Such addition crossover is indeed produced by a 
break of symmetry of the probability distribution function for the velocity of 
the particle leading then to a bias and hence, producing a preferential 
direction of diffusion, yielding in a growth of the average velocity. 
Saturation is again observed for large enough time.

\begin{figure*}[t]
\begin{center}
\centerline{\includegraphics[width=14cm,height=12cm]{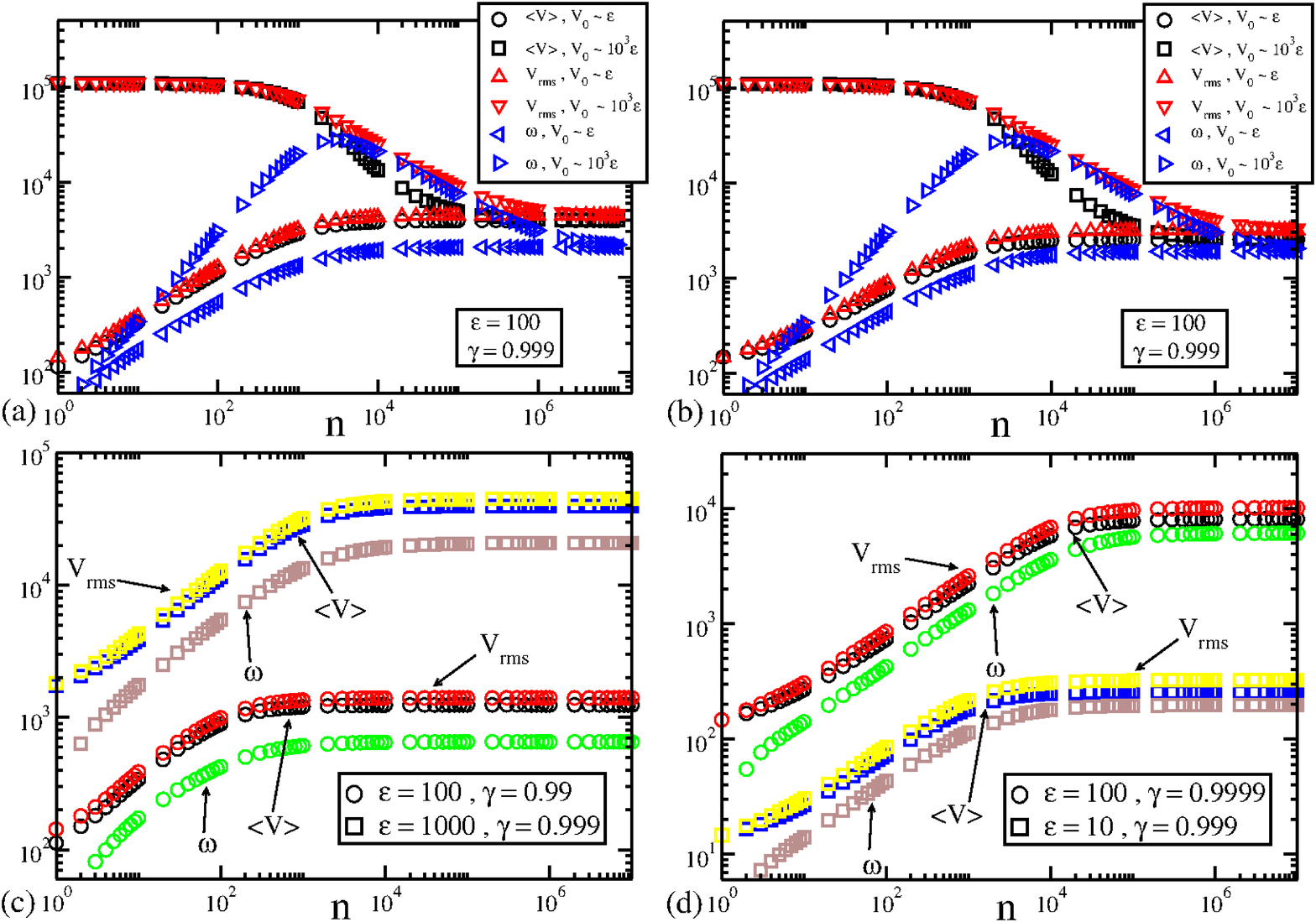}}
\end{center}
\caption{{\it Evolution of $\langle V\rangle$, $V_{rms}$ and $\omega$ as 
function of $n$. The control parameters are shown in the figure. Complete in 
(a) and static wall approximation in (b) show the dynamics considering either 
small and large initial velocities. Small initial conditions are considered in 
(c) and (d), for the complete and static wall approximation. All curves show a  
converge to the stationary state for long times.}}
\label{fig2}  
\end{figure*}

Based on the posed above, we show in Fig. \ref{fig2}, the behavior of $\langle 
V\rangle$ (black circles and squares), $V_{rms}$ (red up and down triangles) 
and 
$\omega$ (blue right and left triangles) as a function of the number of 
collisions $n$. The initial velocities were chosen in two different regimes: 
(i) high\footnote{High as compared to $\epsilon$.} initial velocities 
$(V_0\approx10^3\epsilon)$ and; (ii) low initial velocities 
$(V_0\approx\epsilon)$. We ensemble average the dynamics by considering the 
phase was equally distributed in the range $\phi\in[0,2\pi]$.

A comparison of the saturation of the three observables $\langle V\rangle$, 
$V_{rms}$ and $\omega$ is better seen in Figs. \ref{fig2}(c,d). Important to 
mention is that a change in the parameter $\epsilon$ leads to different 
saturation and it does not affect the crossover time. However, the parameter 
$\gamma$ changes both the saturation (stationary state) and the crossover 
times. 
With a scaling approach, as done previously in the literature, see for instance 
Refs. \cite{ref13,ref14,ref15}, a rescale can be done and overlap both curves, 
of the same observable, into an universal plot. However, in the scenario where
high dissipation is considered, and we have low values for the parameter 
$\epsilon$, 
the scaling invariance is very difficult to be observed, since we have successive
boundary crisis between manifolds and crisis between attractors \cite{crises}.

As we will see in the next section, the numerical values of the saturation 
plateaus play an important role in the Thermodynamics analysis. The values of 
the plateaus for different values of the control parameters are shown in 
Tables \ref{Tab2} and \ref{Tab1}. We see the saturation plateaus for the 
complete version are higher as compared to the static wall approximation. This 
is close connected to the probability distribution function of the velocity in 
the phase space. For short, the particle {\it prefers} to stay with high energy 
in the complete version while compared to the static wall approximation. 
Although the phase space is similar for both versions, their occupation are 
different.

\begin{table}[t]
{
\begin{tabular}{|c|c|c|c|c|c|} \hline \hline
$\epsilon$ & $\gamma$ & $\langle V\rangle$ & $V_{rms}$ & $\omega$
 \\
\hline
$10$&$0.999$&$257.54(5)$&$324.30(5)$&$197.09(2)$\\
\hline
$100$&$0.99$&$793.85(4)$&$995.03(5)$&$599.91(3)$ \\
\hline
$100$&$0.999$&$2531.2(3)$&$3165.9(5)$&$	1901.5(4)$ \\
\hline
$100$&$0.9999$&$8091(9)$&$10079(9)$&$5999(9)$ \\
\hline
$1000$&$0.999$&$25222(4)$&$31611(5) $&$19054(2)$ \\
\hline
\hline
\end{tabular} }
\caption{Simplified mapping: Numerical values for the stationary state for 
$\langle V\rangle$, $V_{rms}$ and $\omega$ considering some pairs of  
($\epsilon$,$\gamma$).}
\label{Tab2}
\end{table}

\begin{table}[t]
{
\begin{tabular}{|c|c|c|c|c|c|} \hline \hline
$\epsilon$ & $\gamma$ & $\langle V\rangle$ & $V_{rms}$ & $\omega$
 \\
\hline
$10$&$0.999$&$407.50(4)$&$461.23(4)$&$216.05(1)$\\
\hline
$100$&	$0.99$&$1244.3(1)$&$1405.6(1)$	&$   653.53(3)$ \\
\hline
$100$&$0.999$&$3959.7(2)$&$4469.0(3)$&$  2071.7(2)$ \\
\hline
$100$&$	0.9999	$&$12736(4)$&$	14333(8)$&$6570(9)$ \\
\hline
$1000$&$0.999$&$39608(1)$&$44694(2)$&$20706(1)$ \\
\hline
\hline
\end{tabular} }
\caption{Complete mapping: Numerical values for the stationary state for 
$\langle V\rangle$, $V_{rms}$ and $\omega$ considering some pairs of  
($\epsilon$,$\gamma$).}
\label{Tab1}
\end{table}

\section{Thermodynamics and discussion}
\label{sec4}
In this section we describe some thermodynamical results for the proposed 
models by an analytical method motivated by Ref. \cite{ref16}. We first present 
our results for the simplified version, see Eq. (\ref{eq4}) and then, latter 
on, for the complete version, written in Eq. (\ref{eq1}).

\subsection{Simplified version}

To describe some of the thermodynamical properties for the simplified model, we 
used the equations of motion (\ref{eq4}) considering many different 
trajectories. We then construct a histogram for the velocity variable, as an 
attempt to have an insight of the probability density function for the velocity.
\begin{figure}[ht!]
\centering
\includegraphics[width=9cm]{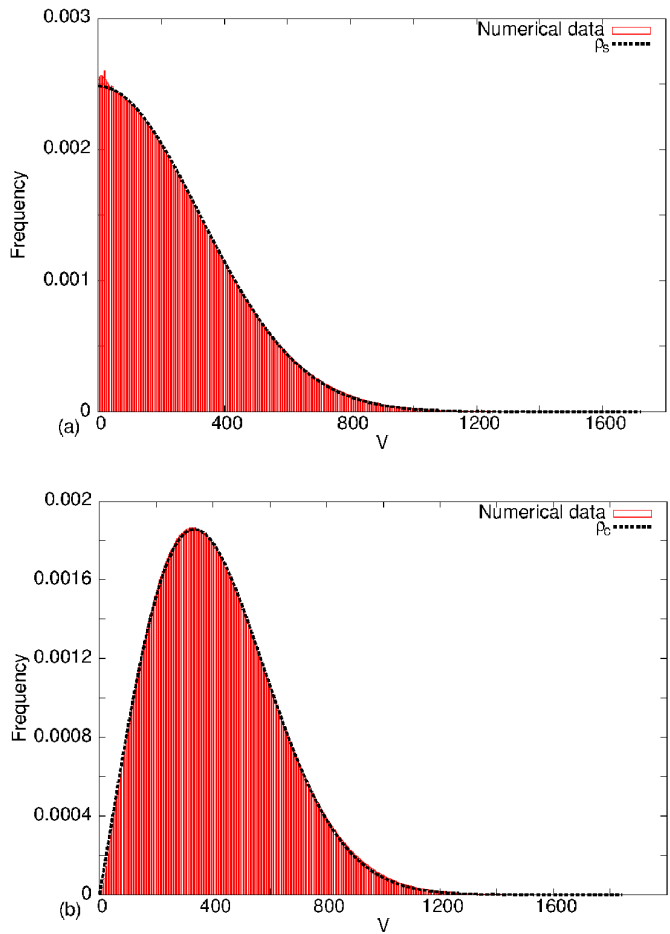}
\caption{\it{Histogram and probability distribution function for: (a) a 
simplified version, (b) complete version of the impact system.}}
\label{fig3}  
\end{figure}  
From Fig. \ref{fig3}(a), we see that the histogram for the velocity has a 
half-Gaussian shape around zero. Such a shape allows us to write the 
probability 
density function for the velocity as a function of the type 
$\rho_{s}(V)=\frac{2}{\sqrt{2\pi}\sigma}e^{-\frac{V^{2}}{2\sigma^{2}}}$ 
for $V\in[0,\infty)$. Also, it can be shown numerically that the distribution probability 
for the phase variable is almost uniform and independent of the velocity variable and the
averages can be taken separately from each other. 
Therefore, the mean squared velocity is given by
\begin{equation}
\langle{V^{2}}\rangle=\int_{0}^{\infty}V^{2}\rho_{s}(V)dV=\sigma^{2}.
\label{eq9*} 
\end{equation}
It is known that for an ideal classical gas the temperature is proportional to 
the mean kinetic energy \cite{ref6}. Hence, we choose $T=\sigma^{2}$ and a 
straightforward integration yields
\begin{equation}
\langle{V}\rangle=\sqrt{\frac{2T}{\pi}}.
\label{eq9} 
\end{equation}

The expression for the temperature can also be obtained directly from the 
mapping (\ref{eq4}). Squaring both sides of the expression for the velocity and 
taking the average over an ensemble of different initial phases 
$\phi\in[0,2\pi]$, we end up with
\begin{equation}
\langle V_{n+1}^{2}\rangle=\gamma^{2}\langle V_{n}^{2}\rangle 
_{V}+\frac{(1+\gamma)^{2}\epsilon^{2}}{2}.
\label{eq9a}
\end{equation}
Here the first term on the right side of the equation is averaged over the 
velocity probability distribution and the second term is obtained after 
averaging over the phase variable. At the stationary state, and considering the 
result of Eq.(\ref{eq9*}), we have $T=\gamma^{2}T+\frac{(1+\gamma)^{2}\epsilon^{2}}{2}$, 
thus yielding
\begin{equation}
T=\frac{(1+\gamma)\epsilon^{2}}{2(1-\gamma)}.
\label{eq10*}
\end{equation}

The other quantities can also be obtained by a similar procedure, as the one done in Eq.(\ref{eq9*}), in particular the root mean square 
velocity
\begin{equation}
\sqrt{\langle{V^2}\rangle}=\sqrt{T}~,
\label{eq10}
\end{equation}
and also the deviation around the mean velocity
\begin{equation}
\omega=\sqrt{\left(1-\frac{2}{\pi}\right)T}~.
\label{eq11}
\end{equation}
Using Eq.(\ref{eq9}), (\ref{eq10}), (\ref{eq11}) and the temperature 
given by equation (\ref{eq10*}) it is possible to recover the same numerical
values for $\langle V\rangle$, $V_{rms}$ and $\omega$ shown in Table (\ref{Tab2}).

\subsection{Complete model}
Let us now move on and discuss the results for the complete model. We proceed in a 
similar way as made to the simplified version. Figure \ref{fig3}(b) shows that 
the probability distribution of $V$ is not described anymore by a semi-Gaussian 
function. It can be approximated by a Weibull distribution \cite{weibull} with 
a shape parameter $k=2$. The probability distribution function is then written 
as $\rho_{c}(V)=\frac{V}{T}e^{-\frac{V^{2}}{2T}}$, and we consider in our 
calculations that $V\in[0,\infty)$. In fairness, the real variation of velocity 
is $[-\epsilon,\infty)$ but the probability of finding a velocity in the 
interval $[-\epsilon,0)$ is very small as compared to the complementary range 
for the parameters considered in this paper. In this case, it can also be shown 
numerically, that the distribution probability for the phase variable is almost
uniform and independent of the velocity variable. From such a distribution, we have
\begin{equation}
\langle{V}\rangle=\sqrt{\frac{\pi T}{2}}.
\label{eq13} 
\end{equation}

To discuss the temperature in terms of the dynamical equations, it turns 
convenient to rewrite the transcendental equation $F(\phi_c)$ in a more 
convenient way as
\begin{equation}
\frac{1}{2}t^{2}-V_{n}t-m\epsilon\cos\left(\phi_{n}
\right)+m\epsilon\cos\left(\phi_{n}+t\right) = 0.
\label{eqG1}
\end{equation}
The parameter $m$ is defined in such a way that for $m=0$ the results for the 
simplified version are obtained. For $m=1$ we consider the complete version 
while for $0<m<1$ the solution for $t$ is required. Suppose $t$ can be 
approximated by
\begin{eqnarray}
t & = & a_{0}+a_{1}m+a_{2}m^{2}+a_{3}m^{3}\ldots\label{eqG1*}
\end{eqnarray}
for $0\leq m\leq1$. Replacing Eq.(\ref{eqG1*}) in the expression (\ref{eqG1}), after some straightforward algebra and rearranging properly the terms, we have
\begin{equation}
\begin{array}{llll}
\left\{ \frac{a_{0}^{2}}{2}-V_{n}a_{0}\right\} +[a_{0}a_{1}-V_{n}a_{1}-\epsilon\cos(\phi_{n})~\\
+\epsilon\cos(\phi_{n}+a_{0})]m+[a_{0}a_{2}-V_{n}a_{2}~\\
+\frac{a_{1}^{2}}{2}-\epsilon\sin(\phi_{n}+a_{0})a_{1}]m^{2}+[a_{0}a_{3}-V_{n}a_{3}~\\
+a_{1}a_{2}-\epsilon\sin(\phi_{n}+a_{0})a_{2}-\frac{\epsilon\cos(\phi_{n}+a_{0})}{2}a_{1}^{2}] m^{3}=0.
\end{array}
\label{eqG2}
\end{equation}

We truncate Eq.(\ref{eqG2}) at the third term and obtain the expressions for $a_{0}[V]$, 
$a_{1}[V,\phi]$, $a_{2}[V,\phi]$ and so on, considering that each element inside 
of the brackets must vanish. First analysis yields $a_{0}=2V$. Because the 
multiple collisions are rare as compared to the whole dynamics, solution of Eq. 
(\ref{eqG1}) is a good approximation to construct the probability. From numerical simulations we know 
that the probability of $V\leq\epsilon$ is small, then the series converges for 
 $0\leq m<1$, hence $\frac{\left|a_{g}\right|}{\left|a_{g+1}\right|}<1$. The 
relations for $a_{0}[V]$, $a_{1}[V,\phi]$, $a_{2}[V,\phi]$ are
\begin{equation}
\left\{\begin{array}{llllll}
a_{0}= 2V_{n},~\\
~\\
a_{1}=\frac{\epsilon}{V_{n}}\left(\cos(\phi_{n})-\cos(\phi_{n}+2V_{n})\right),~\\
~\\
a_{2}=\frac{\epsilon^{2}}{V_{n}^{2}}\left\{\cos(\phi_{n})-\cos(\phi_{n}
+2V_{n})\right\}~\\
+\left\{\sin(\phi_{n}+2V_{n})-\frac{\cos(\phi_{n})-\cos(\phi_{n}+2V_{n})}{2V}
\right\}.
\end{array}
\right.
\label{eqG2*}
\end{equation}

Using Eq.(\ref{eqG1*}) and the expressions given in (\ref{eqG2*}), at the equilibrium state we have

\begin{equation}
\begin{array}{lll}
\langle V_{n+1}\rangle=[\gamma\langle V_{n}\rangle - (1+\gamma)\epsilon\langle 
\sin(\phi_{n}+2V_{n})\rangle]~\\
+[\gamma\langle a_{1}\rangle -(1+\gamma)\epsilon\langle\cos(\phi_{n}+2V_{n})a_{1}\rangle]m~\\
+[\gamma\langle a_{2}\rangle -(1+\gamma)\epsilon\langle 
\cos(\phi_{n}+2V_{n})a_{2}-~\\
\frac{\sin(\phi_{n}+2V_{n})}{2}a_{1}^{2}\rangle] m^{2}
\label{eqG2**}
\end{array}
\end{equation}

The terms $\left\langle \sin\left(\phi_{n}+2V_{n}\right)\right\rangle$ and $\left\langle a_{1}\right\rangle$ 
have zero value after averaging over the phase variable, which is distributed uniformly. Also, one can realize
that $\left\langle \cos\left(\phi_{n}+2V_{n}\right)a_{1}\right\rangle=\langle\frac{\epsilon}{V_{n}}\cos(\phi_{n}+2V_{n})(\cos(\phi_{n})-\cos(\phi_{n}+2V_{n}))\rangle$.
After take an average over the phase, one can obtain
\begin{equation}
\left\langle\cos\left(\phi_{n}+2V_{n}\right)a_{1}\right\rangle=
\left\langle\frac{\epsilon}{V_{n}}\left(\frac{\cos\left(2V_{n}\right)}{2}-\frac{1}{2}\right)\right\rangle_{V}~,
\label{eqaa}
\end{equation}
where the right-hand side term can be expressed by the cosine function expansion as 

\begin{equation}
\left\langle \cos\left(\phi_{n}+2V_{n}\right)a_{1}\right\rangle=
\epsilon\left\langle 
\sum_{l=0}^{\infty}\frac{\left(-1\right)^{l+1}\left(2V_{n}\right)^{2l+1}}{
\Gamma\left(2l+3\right)}\right\rangle _{V}~.
\label{eqaaa}
\end{equation}

The average over the coefficient $\left\langle a_2\right\rangle$ is obtained from
$\langle a_{2}\rangle=\langle\frac{\epsilon^{2}}{V_{n}^{2}}(\cos(\phi_{n})-\cos(\phi_{n}
+2V_{n}))(\sin(\phi_{n}+2V_{n})-\frac{\cos(\phi_{n})-\cos(\phi_{n}+2V_{n})}{2V_{n}})\rangle$.
Considering then, an average over the phase one can obtain 
$\langle a_2 \rangle=\langle \frac{\epsilon^{2}}{V_{n}^{2}}(\frac{\sin(2V_{n})}{2}-\frac{1}{
2V_{n}}(1-\cos2V_{n}))\rangle_{V}$, where now $\langle a_2 \rangle$ is strictly written as function
of the average over the velocity variable. One can expand this last expression for $\langle a_2 \rangle$ 
in power series and obtain
\begin{equation}
\begin{array}{ll}
\langle a_2 \rangle=\langle\frac{\epsilon^{2}}{V_{n}^{2}}[V_{n}+\sum_{l=1}^{\infty}\frac{(-1)^{l}2V_{n})^{2l+1}}{2\Gamma(2l+2)}~\\
-V_{n}-\sum_{l=1}^{\infty}\frac{(-1)^{l}(2V_{n})^{2l+1}}{\Gamma(2l+3)}]\rangle_{V},
\end{array}
\end{equation}
and after rearranging properly the terms, we have
\begin{equation}
\left\langle a_{2}\right\rangle = \epsilon^{2}\left\langle 
\sum_{l=0}^{\infty}\frac{\left(-1\right)^{l+1}2\left(2l+2\right)\left(2V_{n}
\right)^{2l+1}}{\Gamma\left(2l+5\right)}\right\rangle_{V}~.
\label{eqaaaa}
\end{equation}

Finally, the average over the last term of Eq.(\ref{eqG2**}) is given by

\begin{equation}
\left\langle 
\cos\left(\phi_{n}+2V_{n}\right)a_{2}-\frac{\sin\left(\phi_{n}+2V_{n}\right)}{2}a_{1}^{2}\right\rangle=0~.
\label{eqbbbbb}
\end{equation}

For obtainment of Eq.(\ref{eqbbbbb}), we considered that all third order trigonometric functions, 
like $\cos^{3}\left(\phi_{n}\right)$, and their crossed terms, like $\cos\left(\phi_{n}\right)\sin^{2}\left(\phi_{n}\right)$, 
have null averages over the phase variable.

With the previous results obtained in the expressions (\ref{eqaaa}), (\ref{eqaaaa}) and (\ref{eqbbbbb}), one may write Eq.(\ref{eqG2**}) as 
\begin{equation}
\begin{array}{lll}
\left\langle V_{n+1}\right\rangle_{V}=\left\{ \gamma\left\langle 
V_{n}\right\rangle _{V}\right\}+~\\
\left\{ -(1+\gamma)\epsilon^{2}\left\langle 
\sum_{l=0}^{\infty}\frac{\left(-1\right)^{l+1}\left(2V_{n}\right)^{2l+1}}{
\Gamma\left(2l+3\right)}\right\rangle _{V}\right\} m
~\\+\left\{ 
\gamma\epsilon^{2}\left\langle 
\sum_{l=0}^{\infty}\frac{\left(-1\right)^{l+1}2\left(2l+2\right)\left(2V_{n}
\right)^{2l+1}}{\Gamma\left(2l+5\right)}\right\rangle _{V}\right\} m^{2}\label{eqG*2}
\end{array}
\end{equation}

Let us define an auxiliary term $\left\langle V^{2l}\right\rangle _{V}$, then
\begin{eqnarray*}
\left\langle V^{2l}\right\rangle _{V} & = & 
\int_{0}^{\infty}V^{2l}\frac{V}{T}e^{\frac{-V^{2}}{2T}}dV
\end{eqnarray*}
if we call $u=\frac{V}{\sqrt{2T}}$, we have
\begin{eqnarray}
\left\langle V^{2l}\right\rangle _{V} & = & 
\frac{(2T)^{l+1}}{2T}2\int_{0}^{\infty}u^{2(l+1)-1}e^{-u^{2}}du,\nonumber \\
\left\langle V^{2l}\right\rangle _{V} & = & (2T)^{l}\Gamma\left(l+1\right)\label{eqG**2}
\end{eqnarray}
where the $\Gamma$ function is well defined for $l>-1$ \cite{arfken}.

Using Eq.(\ref{eqG**2}) the expression of the average velocity, Eq. (\ref{eqG*2}), can then be written as

\begin{equation}
\begin{array}{lll}
\left\langle V_{n+1}\right\rangle _{V} = \left\{ \gamma\left\langle 
V_{n}\right\rangle _{V}\right\}+~\\
\left\{ -(1+\gamma)\epsilon^{2}\sum_{l=0}^{\infty}\frac{\left(-1\right)^{l+1}2^{2l+1}
\left(2T\right)^{l+1/2}\Gamma\left(l+3/2\right)}{\Gamma\left(2l+3\right)}
\right\} m+~\\
\left\{ 
\gamma\epsilon^{2}\sum_{l=0}^{\infty}\frac{\left(-1\right)^{l+1}
2\left(2l+2\right)2^{2l+1}\left(2T\right)^{l+1/2}\Gamma\left(l+3/2\right)}{
\Gamma\left(2l+5\right)}\right\} m^{2},
\end{array}
\end{equation}
after rearranging properly the terms

\begin{equation}
\begin{array}{lll}
\left\langle V_{n+1}\right\rangle _{V}=\gamma\left\langle 
V_{n}\right\rangle_{V}+~\\
(1+\gamma)\epsilon^{2}\left(8T\right)^{1/2}\left\{ 
\sum_{l=0}^{\infty}\frac{\left(-8T\right)^{l}\Gamma\left(l+3/2\right)}{
\Gamma\left(2l+3\right)}\right\}m-~\\
2\gamma\epsilon^{2}\left(8T\right)^{1/2}\left\{ 
\sum_{l=0}^{\infty}\frac{\left(2l+2\right)\left(-8T\right)^{l}
\Gamma\left(l+3/2\right)}{\Gamma\left(2l+5\right)}\right\} m^{2}
\label{eqG3}
\end{array}
\end{equation}

Recalling the following mathematical relation for the gamma function \cite{arfken}.
\begin{eqnarray}
\Gamma\left(2z\right) & = & 
\left(\pi\right)^{-\frac{1}{2}}2^{2z-1}\Gamma\left(z\right)\Gamma\left(z+\frac{1
}{2}\right),
\end{eqnarray}
we may obtain after some straightforward algebra
\begin{eqnarray}
\sum_{l=0}^{\infty}\frac{\left(-8T\right)^{l}\Gamma\left(l+3/2\right)}{
\Gamma\left(2l+3\right)} & = & 
\frac{\sqrt{\pi}}{4}\sum_{l=0}^{\infty}\frac{\left(-2T\right)^{l}}{
\Gamma\left(l+2\right)},\nonumber \\
 & = & \frac{\sqrt{\pi}}{8T}\left(1-e^{-2T}\right)\label{eqG4}.
\end{eqnarray}

The last term of Eq. (\ref{eqG3}) stays as $\sum_{l=0}^{\infty}\frac{\left(2l+2\right)\left(-8T\right)^{l}
\Gamma\left(l+3/2\right)}{\Gamma\left(2l+5\right)}=  
\frac{\sqrt{\pi}}{4}\sum_{l=0}^{\infty}\frac{\left(-2T\right)^{l}}{\left(l+2\right)\left(2l+3\right)\Gamma\left(l+1\right)}$. Again, 
rearranging the terms we have

\begin{equation}
\begin{array}{ll}
\sum_{l=0}^{\infty}\frac{\left(2l+2\right)\left(-8T\right)^{l}
\Gamma\left(l+3/2\right)}{\Gamma\left(2l+5\right)}=~\\
\frac{\sqrt{\pi}}{4}\sum_{l=0}^{\infty}\left[-\frac{\left(-2T\right)^{l}}{
\left(l+2\right)\Gamma\left(l+1\right)}+\frac{\left(-2T\right)^{l}}{
\left(l+3/2\right)\Gamma\left(l+1\right)}\right]
\label{eqG5}
\end{array}
\end{equation}

Now we proceed to evaluate the sums in Eq.(\ref{eqG5}) with the following steps \cite{dingle}: 
First we use the fact that $\frac{1}{n+1}=\int_{0}^{1}u^{n}du$, obtaining thus
\begin{equation}
\begin{array}{ll}
 \sum_{l=0}^{\infty}\frac{\left(2l+2\right)\left(-8T\right)^{l}
\Gamma\left(l+3/2\right)}{\Gamma\left(2l+5\right)}=~\\ 
\frac{\sqrt{\pi}}{4}\sum_{l=0}^{\infty}\left[-\frac{\left(-2T\right)^{l}}{
\Gamma\left(l+1\right)}\int_{0}^{1}u^{l+1}du+\frac{\left(-2T\right)^{l}}{
\Gamma\left(l+1\right)}\int_{0}^{1}u^{l+\frac{1}{2}}du\right]~,
\end{array}
\end{equation}

\begin{equation}
\begin{array}{ll}
 \sum_{l=0}^{\infty}\frac{\left(2l+2\right)\left(-8T\right)^{l}
\Gamma\left(l+3/2\right)}{\Gamma\left(2l+5\right)}=~\\
\frac{\sqrt{\pi}}{4}\left[-\int_{0}^{1}e^{-2Tu}udu+\int_{0}^{1}e^{-2Tu}u^{\frac{
1}{2}}du\right]~,
\end{array}
\end{equation}
then we interchange the order of the summation and the integration. After that, we perform the sum over $l$, finally we integrate.

\begin{equation}
\begin{array}{ll}
 \sum_{l=0}^{\infty}\frac{\left(2l+2\right)\left(-8T\right)^{l}
\Gamma\left(l+3/2\right)}{\Gamma\left(2l+5\right)}=~\\
\frac{\sqrt{\pi}}{4}\left[-\frac{1}{\left(2T\right)^{2}}+\frac{e^{-2T}}{
\left(2T\right)^{2}}+\frac{\sqrt{\pi}}{2\left(2T\right)^{\frac{3}{2}}}
erf\left(\sqrt{2T}\right)\right]~,
\label{eqG6}
\end{array}
\end{equation}
where $erf\left(x\right)=\frac{2}{\sqrt{\pi}}\intop_{0}^{x}e^{-x^{2}}dx$
is the error function and is in agreement with 
$\lim_{x\rightarrow\infty}erf\left(x\right)=1$. Therefore, for high 
temperatures, after replacing Eqs.(\ref{eqG4}) and (\ref{eqG6}) in Eq.(\ref{eqG3}), making $m=1$ and putting $\left\langle V\right\rangle$ in evidence we end up with
\begin{equation}
\left\langle V\right\rangle = 
\frac{1}{1-\gamma}\left[\frac{(1+\gamma)\epsilon^{2}}{2}\sqrt{\frac{\pi}{2T}}
+\gamma\epsilon^{2}\frac{\pi}{\left(4T\right)}\right].
\end{equation}
The first term on the right does indeed contributes at the limit of high 
temperatures, then, using Eq. (\ref{eq13}) we find that
\begin{equation}
T = \frac{(1+\gamma)\epsilon^{2}}{2\left(1-\gamma\right)},
\label{eq14*}
\end{equation}
which is in remarkable well agreement with the result obtained for the 
simplified version of the model obtained in Eq. (\ref{eq10*}). Similar to discussed for the simplified 
version, we found also
\begin{equation}
\sqrt{\langle{V^2}\rangle}=\sqrt{2T}~,
\label{eq14}
\end{equation}
and the deviation around the average velocity
\begin{equation}
\omega=\sqrt{\left(2-\frac{\pi}{2}\right)T}~.
\label{eq15}
\end{equation}
Using equations (\ref{eq13}), (\ref{eq14}), (\ref{eq15}) and the temperature
given by equation (\ref{eq14*}) it is possible to recover the same numerical values for $\langle V\rangle$, $V_{rms}$ and $\omega$ shown in Table (\ref{Tab1}).

\subsection{Discussion}

Our findings shown in the previous sections were obtained from different approaches: (i) via numerical 
simulations; (ii) by the use direct average of the equation of the velocity; 
and (iii) by the probability distribution of the velocity. The agreement 
between these three approaches is remarkable. Let us now obtain a relation between $\langle{V}\rangle$, $\sqrt{\langle{V^2}\rangle}$, and $\omega$. For 
that we define new variables as $X=\ln(\langle{V}\rangle)$, 
$Y=\ln(\sqrt{\langle{V^2}\rangle})$, $Z=\ln(\omega)$. For the simplified 
version we obtain the following relations from Eqs. (\ref{eq9}), (\ref{eq10}) and (\ref{eq11})
\begin{equation}
Y=X+\frac{1}{2}\ln\left(\frac{\pi}{2}\right)~,
\label{eq16} 
\end{equation}
\begin{equation}
Z=X+\frac{1}{2}\ln\left(\frac{\pi}{2}-1\right)~,
\label{eq17}
\end{equation}
\begin{equation}
Z=Y+\frac{1}{2}\ln\left(1-\frac{2}{\pi}\right)~.
\label{eq18}
\end{equation}

\begin{figure*}[ht!]
\centering
\includegraphics[width=17cm,height=10.0cm]{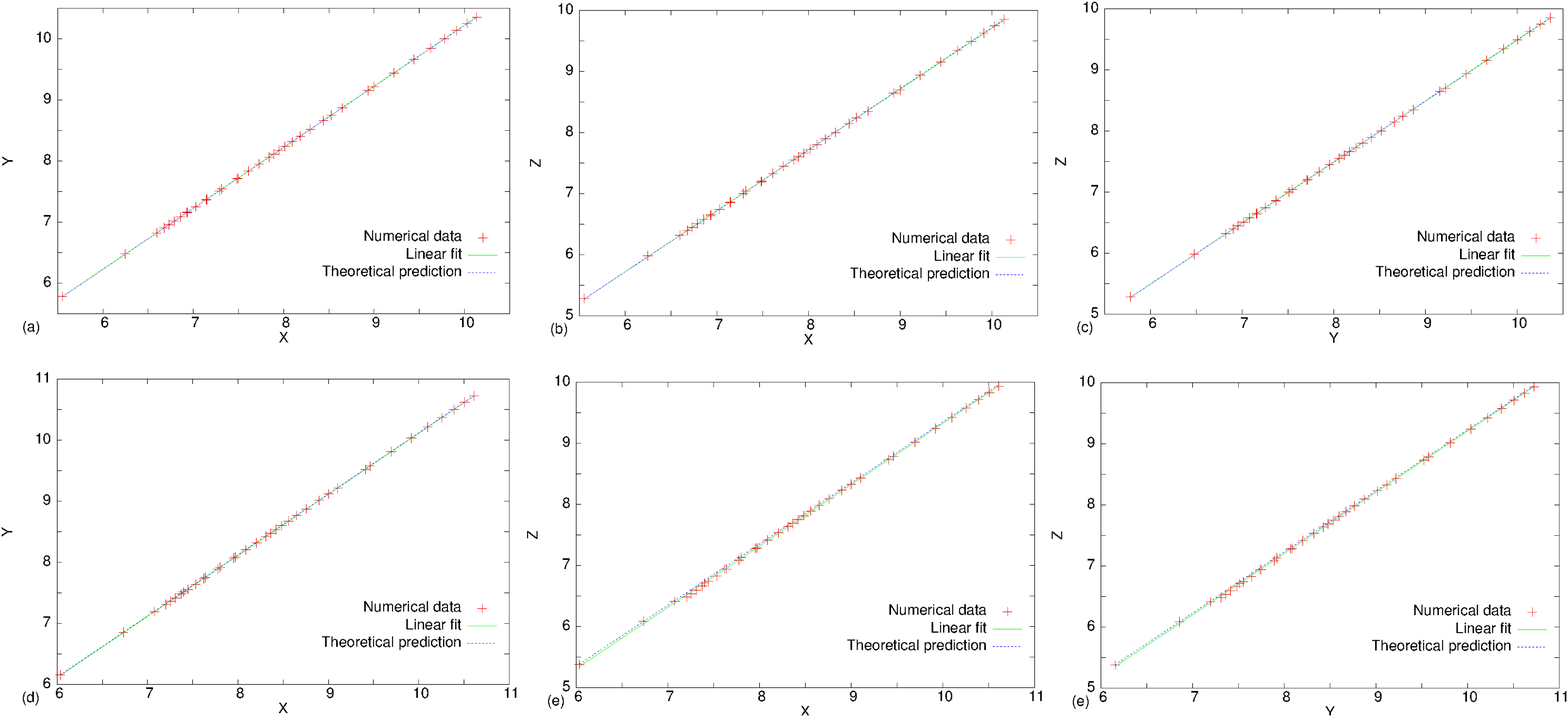}
\caption{\it{Plot of the observables obtained from numerical data, linear fit 
and theoretical prediction. Simplified version is shown in (a) for Eq. 
(\ref{eq16}), (b) for Eq. (\ref{eq17}), (c) for Eq. (\ref{eq18}), while 
complete version is shown in (d) for Eq. (\ref{eq19}), (e) for Eq. (\ref{eq20}) 
and, (f) for Eq. (\ref{eq21}).}}
\label{fig4}  
\end{figure*}

For the complete version, the relations from Eqs. (\ref{eq13}), (\ref{eq14}) and (\ref{eq15}) are
\begin{equation}
Y=X+\frac{1}{2}\ln\left(\frac{4}{\pi}\right)~,
\label{eq19} 
\end{equation}
\begin{equation}
Z=X+\frac{1}{2}\ln\left(\frac{4}{\pi}-1\right)~,
\label{eq20}
\end{equation}
\begin{equation}
Z=Y+\frac{1}{2}\ln\left(1-\frac{\pi}{4}\right)~.
\label{eq21}
\end{equation}

The behavior shown in Fig. \ref{fig4}, the comportment of equations 
(\ref{eq16}-\ref{eq21}) and numerical data regarding both the simplified and complete model, shows a remarkable agreement between the theory developed in this paper  and the 
numerical results. 

To illustrate better the novelty and results obtained in this paper we shown Table (\ref{Tab3})
 which contains a comparison for the $\omega$ variable regarding analytical results, from Eqs.(\ref{eq11})
 and (\ref{eq10*}) for the simplified model (ASM),  and Eqs.(\ref{eq15}) and (\ref{eq14*}) for the complete model (ACM),
 with the numerical findings (NSM) and (NCM) respectively, shown in Tables (\ref{Tab2}) and (\ref{Tab1}). One can see that the agreement is
 quite good, which gives robustness to the theory developed in this study. Besides, it opens the possibility for the formalism
 to be extended to other similar dynamical systems, including billiard problems.

\begin{table}[t]
{
\begin{tabular}{|c|c|c|c|c|c|c|} \hline \hline
$\epsilon$ & $\gamma$ & $\omega_{ASM}$ & $\omega_{NSM}$ & $\omega_{ACM}$ & $\omega_{NCM}$
 \\
\hline
$10$&$0.999$&$190.58$&$197.09(2)$&$207.12$&$216.05(1)$\\
\hline
$100$&$0.99$&$601.30$&$599.91(3)$&$653.50$&$653.53(3)$ \\
\hline
$100$&$0.999$&$1905.8$&$1901.5(4)$&$2071.2$&$2071.7(2)$ \\
\hline
$100$&$0.9999$&$6028$&$5999(9)$ &$6551$&$6570(9)$\\
\hline
$1000$&$0.999$&$19058$&$19054(2)$&$20712$&$20706(1)$ \\
\hline
\hline
\end{tabular} }
\caption{ comparison for the $\omega$ variable regarding analytical results for the simplified model (ASM) and for 
the complete model (ACM), with the numerical findings for the simplified approach (NSM) and the complete one (NCM). }
\label{Tab3}
\end{table}

\section{Final Remarks and Conclusions}
\label{sec5}

The dynamics of a dissipative impact system was described by nonlinear mappings 
for two different versions, complete and simplified, for the velocity of the 
particle and the phase of the vibrating wall. Dissipation was introduced via 
inelastic collisions leading the existence of attractors in the phase space.

A numerical and statistical 
investigation for the variables $\langle V\rangle$, $V_{rms}$ and $\omega$ 
(deviation of the average velocity) was made for both versions of the mappings. 
For long time series, these observables bend towards a saturation plateau 
which marks the stationary state. Such a regime varies as the control 
parameters associated with the dissipation $(\gamma)$ and ratio between 
acceleration $(\epsilon)$ are changed.

At the stationary state, the square velocity can be obtained. From equipartition 
theorem, such observable can be interpreted as an equilibrium temperature 
\cite{ref16}. We obtained analytical equations for the  $\langle V\rangle$, $V_{rms}$ and $\omega$ variables in the equilibrium state as functions of the
 parameters of the model, with these equations we were able to calculate 
the numerical values of those variables without doing the simulations . 
A remarkable 
assembly was obtained considering both numerical 
and theoretical investigation, between statistical and thermal variables. This 
result gives robustness to the formalism, and opens 'new doors' for similar 
analysis in other more complex dynamical systems, particularly in time 
dependent billiards.

\section*{Acknowledgements}
GDI thanks to the Brazilian agency CAPES. ALPL acknowledges FAPESP 
(2014/25316-3) and CNPq for financial support. EDL kindly acknowledges support 
from CNPq (303707/2015-1), FAPESP (2012/23688-5) and FUNDUNESP.



\end{document}